\documentclass[twocolumn,nofootinbib,amsmath,prl,aps,superscriptaddress,tightenlines,preprintnumbers]{revtex4}
\usepackage{graphicx}

\def\bea{\begin{eqnarray}}
\def\eea{\end{eqnarray}}

\begin{document}

\newcount\hour \newcount\minute
\hour=\time \divide \hour by 60
\minute=\time
\count99=\hour \multiply \count99 by -60 \advance \minute by \count99
\newcommand{\mydate}{\ \today \ - \number\hour :00}

\title{Electroweak Sudakov Corrections and the Top Quark Forward-Backward Asymmetry }

\author{Aneesh V.\ Manohar}
\affiliation{Department of Physics, University of California at San Diego, 
  9500 Gilman Drive, La Jolla, CA 92093}

\author{Michael Trott}
\affiliation{Theory Division, Physics Department, CERN, CH-1211 Geneva 23, Switzerland}

\begin{abstract}
The Standard Model (SM) prediction of the top quark forward backward asymmetry is shown to be slightly enhanced by a correction factor of $1.05$ due to electroweak Sudakov (EWS) logarithms of the form  $(\alpha/\sin^2 \theta_W)^n \, \log^{m\le 2n} (s/M_{W,Z}^2)$. The EWS effect on the dijet and $t \overline t$ invariant mass spectra is significant, reducing the SM prediction by $ \sim 20, 10 \%$ respectively for the highest invariant masses measured at the LHC, and changing the shape of the high-mass tail of the spectrum. These corrections 
significantly affect measurements of the top quark invariant mass
spectrum and the search for an excess of events related to $A_{\rm FB}^{t \bar{t}}$. 
\end{abstract}

\maketitle
\newpage
 {\bf I. Introduction:} In high-energy scattering processes at the LHC, with partonic center-of-mass energies $\sqrt{\hat s}$ much larger than the electroweak scale, the $W$ and $Z$ bosons act as massless particles in an unbroken gauge theory. The wide separation of scales, $s \gg M_{W,Z}^2$ leads to Sudakov double logarithms $\alpha_W L^2$, $L= \log s/M_{Z,W}^2$, $\alpha_W=\alpha/\sin^2 \theta_W$, at each order in perturbation theory, which can be substantial (e.g.\  $\sim\! 37$\% for $WW$ production at 2~TeV). While QCD Sudakov corrections cancel for inclusive processes, the electroweak ones do not, because the incident beams are not electroweak singlets~\cite{ciafaloni}. Recently~\cite{Chiu} effective field theory (EFT) methods have been used to systematically sum the electroweak Sudakov (EWS) corrections using renormalization group methods. The EFT result is naturally given in terms of $\log \sigma$, and has the schematic form
\begin{eqnarray}
\log \sigma &=& L f_0(\alpha_W L) + f_1 (\alpha_W L) + \ldots
\end{eqnarray}
in terms of the leading-log series $f_0$, the next-to-leading-log series $f_1$, etc. The EFT computation neglects power corrections of the form $M_Z^2/\hat s$, but includes the complete dependence on electroweak scale mass-ratios such as $M_W/M_Z$, $m_t/M_Z$ and $M_H/M_Z$. The results in Ref.~\cite{Chiu} include the complete NLL series including Higgs effects, and the most important terms in the NNLL series. The omitted NNLL corrections are Higgs effects in the three-loop cusp anomalous dimension and a two-loop matching correction which are not known. The EWS resummation can not be performed simply by exponentiating a fixed order result, because there are different color and flavor structures which mix under  renormalization group evolution in the EFT.
While the EFT is formally not valid near threshold, numerically, the results are still quite accurate because the EWS corrections are not log-enhanced in this region (see Ref.~\cite{Chiu}).
 
The EWS logs grow with energy, and are  important for large invariant mass measurements, such as the recent CDF measurement
of the top-quark forward-backward asymmetry, $A_{\rm FB}^{t \bar{t}}(M_{t  \bar{t}})$~\cite{Aaltonen:2011kc}, which has a $\sim \! 3 \sigma$ deviation from the SM prediction~\cite{Kuhn:2011ri} at $M_{t \bar{t}}>450\,{\rm GeV}$.
In this paper, we study EWS effects on observables needed to study $A_{\rm FB}^{t \bar{t}}(M_{t  \bar{t}})$, as a function of the $t \bar t$ invariant mass $M_{t  \bar{t}}$.  Since the EWS effects are a multiplicative correction, we present them as rescaling factors, by taking the ratio of $A_{\rm FB}^{t \bar{t}}$ computed with and without the EWS effect. This greatly reduces the sensitivity of our results to the choice of parton distribution functions (PDF) or QCD corrections.

We find that the SM EWS corrections enhance $A_{\rm FB}^{t \bar{t}}$ by a factor $1.05$.\footnote{ 
The correction can be applied to partonic calculations (even those including non SM interactions) when calculating
if the process has the same $\rm SU(2) \times U(1)$ gauge flow as the SM. Note, however, that our final results are given averaging over quark spins, and SM corrections are different for left and right-handed quarks.
We restrict ourself to $0.1\, \pi \le \theta_{\rm CM}\le 0.9\,\pi$ to avoid soft scattering. This cut is less restrictive than experimental cuts.} They also suppress the
$d \sigma/ d M_{t \bar{t}}$ spectrum at large invariant mass,  which is
crucial in attempts to understand if the $A_{\rm FB}^{t \bar{t}}$ anomaly is a sign of new physics or not.
We emphasize that 
 \emph{neglected SM electroweak Sudakov corrections can cancel a $t$-channel driven rise in $d \sigma/ d M_{t \bar{t}}$  of $\sim\!\! 10\%$ at large $M_{t \bar{t}}$ due to possible new physics associated with the $A_{\rm FB}^{t  \bar{t}}$ anomaly}. 
 
The overall effect of EWS effects on phenomenology related to $A_{\rm FB}^{t \bar{t}}$ can be even more significant when they also impact attempts to measure the top quark invariant mass spectrum indirectly.
For example, in highly boosted top studies~\cite{boosted} a precise understanding of the normalization and shape of the SM dijet invariant mass spectrum
is essential, and we will show there is also a suppression of  $\sim \! \!10-20 \%$ due to EWS logs for large invariant mass dijet events. 
These SM effects are unaccounted for in current Monte Carlo tools.\footnote{Herwig and Sherpa include pure QED soft and collinear photon re-summations~\cite{Gleisberg:2008ta}, but not these EWS corrections.} 

{\bf II. $t \bar{t}, b \bar{b}, c \bar{c}$ {\bf Phenomenology:}}\label{ttbar}
In~\cite{Kuhn:1998kw} the LO $\rm SM$ asymmetry $A_{\rm FB}^{t  \bar{t}}$ was computed from the $\mathcal{O}(\alpha_s^3)$  cross-section, and a subset of the fixed order $\mathcal{O}(\alpha \, \alpha_s^2)$ terms were also determined. These calculations are based on earlier results~\cite{Berends:1973fd} on the $e^+ e^- \to \mu^+ \mu^- \gamma$ asymmetry, and $ q \overline q \to Q \overline Q g$~\cite{Halzen:1987xd}. A recent SM calculation~\cite{Hollik:2011ps} extended the calculation of $\mathcal{O}(\alpha \, \alpha_s^2)$ terms
and included $\mathcal{O}(\alpha^2)$ corrections from photon radiation. The effect of next-to-leading as well as next-to-next-to leading logarithmic QCD
corrections have been studied in~\cite{Almeida:2008ug}. An interesting discrepancy remains between the SM prediction of $A_{\rm FB}^{t  \bar{t}}$ at large
invariant mass ($M_{t \bar{t}}>450\,{\rm GeV}$), and the CDF measurement~\cite{Aaltonen:2011kc}. 

The  EFT method we use can be illustrated using the process $q \bar q \to t \bar t$ for left-handed quarks. At the high scale $\mu=Q=\sqrt{\hat s}$, the scattering is given by an effective Lagrangian
\begin{eqnarray}
L &=& C_{11} \, q t^a T^A \overline q \ Q t^a T^A \overline Q + C_{12}\, q t^a \overline q\ Q t^a\overline Q \nonumber\\
&&+ C_{21}\, q T^A\overline q \ QT^A \overline Q
+C_{22}\, q \overline q\ Q \overline Q
\label{2}
\end{eqnarray}
where $q=(u,d)_L$ or $(c,s)_L$ are light quark doublets, and $Q=(t,b)_L$ is the heavy quark doublet. $T^A$ are color matrices,  $t^a$ are $SU(2)$ matrices
and $C_{ij}(\mu)$ are hard-matching coefficients. At tree-level, $c(Q)$ is given by single gauge boson exchange. Gluon exchange gives $C_{21}=4 \pi \alpha_s/Q^2$, $W$ exchange gives $C_{12}=4 \pi \alpha_W/Q^2$, and $B$ exchange gives $C_{22}=4 \pi \alpha/\cos^2 \theta_W (1/6)^2$.
At one-loop, $C_{ij}(\mu)$ are given by computing the finite part of one-loop graphs such as box-graphs with all low scales such as $M_Z$ set to zero. The hard-matching $C_{ij}(\mu)$ is computed at the scale $\mu=Q$, and does not contain any large logarithms. The Lagrangian is evolved in the EFT to a low-scale of order $M_Z$, and then the scattering cross-section is taken by squaring the EFT amplitude and integrating with PDFs. The EWS terms arise from the renormalization group evolution of the coefficients $C_{ij}$ from $\mu=Q$ down to $\mu\sim M_Z$. This method has been checked against fixed order computations up to two-loop order, and details can be found in Ref.~\cite{Chiu}.

Here we report on the numerical computation of EWS corrections to dijet and $t \overline t$ production. These corrections are defined as
\bea
\mathcal{R}_{\rm FB}(t) =  \frac{\sigma_{\rm  FB}^{\rm QCD+EWS}(t \bar{t})}{\sigma_{\rm FB}^{\rm QCD}(t \bar{t})}, \quad \mathcal{R}_t =  \frac{\sigma_{t  \bar{t}}^{\rm QCD+EW}}{\sigma_{t  \bar{t}}^{\rm QCD}}\,.
\label{3}
\eea
$\sigma_{\rm FB}$ and $\sigma_{t \bar t}$ are the forward-backward asymmetry and the total cross-section. The superscript $\rm QCD+EW$ means that the EFT calculation is done using the full standard model, and $\rm QCD$ means that QCD alone has been used. The cross-sections include virtual electroweak effects, but not real radiation of additional EW bosons. In dijet production, for example, such events would be part of the $W,Z+$jets signal. With this definition,
multiplying by $\mathcal{R}$ converts a QCD computation into one including EWS corrections as well. The QCD computation can be done using an EFT, or by any other method. The ratios are  insensitive to the choice of PDF.
\begin{table}
\setlength{\tabcolsep}{5pt}
\center
\begin{tabular}{c|c|c|c|c|c|c} 
\hline \hline 
Bin [GeV]&  \multicolumn{3}{c|}{$A_{\rm FB}^{t  \bar{t}} (\%)$} &$\mathcal{R}_{\rm FB}(t)$  &$\mathcal{R}^{\alpha_s^2}_{\rm FB}(t)$ & $\mathcal{R}_t$  
\\
\hline
$[2 \, m_{t\bar t}, 1960]$ & 7.7 & 7.5 & 1.6 & 1.02 & 1.03 & 0.98 
\\
$[2 \, m_{t\bar t}, 450]$ & 5.6 & 5.4 & $-$ & 1.02 & 1.03 & 0.98 %
\\
$[450, 900]$ & 11 & 12 & 2.3 & 1.02 &1.04 & 0.97 %
\\
\hline \hline
\end{tabular}
\caption{The EWS corrections for the Tevatron. The second column gives $A_{FB}^{t  \bar{t}}$ for our SM QCD calculation applying  the  EWS correction,
the third column applies the EWS correction to the quoted central value of the QCD NLO +NNLL calculation of Ahrens {\it et al.} \cite{Almeida:2008ug}. The fourth column
quotes the contribution due to the fixed order EW terms of~\cite{Hollik:2011ps}. There is overlap between our EWS calculation and the results of Ref.~\cite{Hollik:2011ps}. We estimate this double counting is $\sim 0.5\%$ in $A_{FB}^{t  \bar{t}}$. With this caveat, columns three and four can be added.}
\label{table:tcorrections} \vspace{-0.25cm}
\end{table}

We incorporate  EWS corrections by modifying the analytic results of~\cite{Kuhn:1998kw} using the results of Ref.~\cite{Chiu}. The asymmetry is defined as the ratio $\mathcal{A}=(F-B)/(F+B)$, where $F$ and $B$ are the cross-section in the forward and backward hemisphere. In QCD, $F$ and $B$ are order $\alpha_s^2$, whereas $F-B$ is  order $\alpha_s^3$ because the order $\alpha_s^2$ cross-section is $FB$ symmetric. The EWS corrections are \emph{not} $FB$ symmetric.
There are three contributions that we include that have been previously neglected: (a) the change in the normalization of the LO cross section of order $\alpha_s^2 \, \alpha_{W}^n L^{m \le 2n}$ given by $\mathcal{R}_t$ which multiplies the denominator in $\mathcal{A}$.
(b) a new term in the numerator of $\mathcal{A}$ of order $\alpha_s^2 \, \alpha_{W}^n L^{m \le 2n}$ from multiplying the $FB$ symmetric QCD cross-section by the EWS corrections ($\mathcal{R}_{\rm FB}^{\alpha_s^2}(t)$ in Table~\ref{table:tcorrections}). (c) the effect of EWS corrections on the leading QCD FB asymmetry of order $\alpha_s^3 \, \alpha_{W}^nL^{m \le 2n}$. The sum of (b) and (c) is $\mathcal{R}_{\rm FB}(t)$ in Eq.~(\ref{3}), and is the total rescaling of the numerator of $\mathcal{A}$.\footnote{There is also an induced contribution of order $\alpha_s^3 \, \alpha_{W}^n L^{m \le 2n}$ from the $\alpha_s^3$ $FB$ symmetric cross-section, which is smaller than (b), and has been neglected. We have
also neglected the flavor excitation process $q g \to q t \overline t$ as it is highly suppressed~\cite{Kuhn:1998kw}.} 

We use NLO MSTW PDFs~\cite{Martin:2009iq} with the LO QCD results and $\mu = m_t = 173.1\, {\rm GeV}$ for the factorization and renormalization scales.
$\alpha_s$ is set by the MSTW fit value: $\alpha_s(M_Z) = 0.12018$. Numerical values are given in Table~\ref{table:tcorrections}.
We find $A_{\rm FB}^{t  \bar{t}} = 7.4 \%$ and $A_{\rm FB}^{t \bar t}(m_{t\bar t}<450) = 5.3 \%$,  $A_{\rm FB}^{t\bar t}(m_{t\bar t}>450) = 10.7 \%$
for the purely QCD asymmetry, in good agreement with other determinations~\cite{Kuhn:1998kw,Almeida:2008ug,Hollik:2011ps,Kuhn:2011ri}. We find that (a) and (c) essentially cancel.\footnote{Cancelations of  some EWS corrections in $A_{\rm FB}$, were also noted in~\cite{Kuhn:2001hz}, which studied an $\rm SU(2)$ theory.}
The overall rescaling of $\mathcal{A}$ is $\mathcal{R}_{\rm FB}(t)/\mathcal{R}_t$. In all the mass bins that we have considered, the net EWS effect is an enhancement of $A_{\rm FB}^{t \bar{t}}$ by a factor of $1.05$.
 
There is some interest in measuring $A_{\rm FB}^{b  \bar{b}}$, $A_{\rm FB}^{c  \bar{c}}$ to investigate the possible origin of
the $A_{\rm FB}^{t  \bar{t}}$ anomaly~\cite{Bai:2011ed,Strassler:2011vr,Kahawala:2011sm}. The EWS corrections for these observables are given in Table \ref{table:bcorrections}. 
\begin{table}
\setlength{\tabcolsep}{3pt}
\center \vspace{-0.05cm}
\begin{tabular}{c|c|c|c|c|c|c|c|c } 
\hline \hline 
Bin$[\rm GeV]$& \multicolumn{2}{c|}{$A_{\rm \rm FB}^{b \bar{b}}$ (\%)} & $\mathcal{R}_{\rm FB}(b)$  & $\mathcal{R}_b$& 
\multicolumn{2}{c|}{$A_{\rm FB}^{c \bar{c}}$ (\%)}  & $\mathcal{R}_{\rm FB}(c)$  & $\mathcal{R}_c$ \\
\hline
$[50,1960]$ &  0.4 &0.4  & 1.06 & 0.99 & 0.3  &0.3  & 0.99 & 0.99 \\
$[50,350]$ & 0.4  & 0.4  & 1.06 & 0.99 & 0.3  &0.3  & 0.98 & 0.99 \\
$[350,650]$ &  8.1  &7.8  & 1.00 & 1.00 & 6.7  &6.6 & 1.04 & 1.00 \\
$[650,950]$ &  20  &17  & 0.97 & 0.98 & 18  &16  & 1.06 & 0.99 \\
\hline \hline
\end{tabular}
\caption{The EWS corrections and  the uncorrected asymmetry $A_{\rm FB}^{b\bar b}$, $A_{\rm FB}^{c \bar c}$  at the Tevatron.
The left and right columns  are with renormalization and factorization scale $\mu=M_Z$ and $\mu = \sqrt{\hat s}$,
respectively. The  EWS correction is very weakly dependent on the scale choice. In this table $\mathcal{A}^{q \bar q}_{\rm FB}$ is
the uncorrected asymmetry (unlike Table~\ref{table:tcorrections}).
The EWS corrected asymmetry is $\mathcal{A}^{q \bar q}_{\rm FB} \mathcal{R}_{\rm FB}(q)/\mathcal{R}_q$ for $q=b,c$.}
\label{table:bcorrections} \vspace{-0.4cm}
\end{table}
We have normalized the $A_{\rm FB}$ calculations by the LO QCD cross section, as
no complete NLO correction of the asymmetric cross section is known. This approach leads to the estimate
of $A_{\rm FB}^{t  \bar{t}}$ being larger (by about $1.3$) than the results when normalized by the NLO cross section, such as with MCFM~\cite{Kuhn:2011ri}.
Normalizing by inclusive NLO cross sections $\sigma_{f \bar{f} X}$ ($f = b,c$) will lead to an even larger reduction for $A_{\rm FB}^{f  \bar{f}}$,
due to the $t$-channel singularity enhancing  production via $gg \rightarrow gg \rightarrow f  \bar{f}  X$. 
For this, and other reasons ~\cite{Bai:2011ed,Strassler:2011vr,Kahawala:2011sm}, the reported asymmetries are extremely challenging to probe experimentally.

EWS corrections only make a small change to the total cross section, since $\sigma$ is dominated by low invariant mass events because of the PDFs, where the EWS correction is small. However, the tails of the invariant mass distributions have significant EWS corrections that grow in importance with invariant mass. For the $t \overline t$ mass bins reported by CDF~\cite{cdfbin}, the correction factors $\mathcal{R}_t$ are $\{0.99,0.98,0.98,0.97,0.97,0.96,0.96,0.95\}$. The $\mathcal{R}_{b,c}$ corrections are less than 2\% in this region. At the LHC, there are larger effects due
to EWS corrections. Some values of $\mathcal{R}_{t,b,c,}$ are given in Table~\ref{table:tcorrectionsLHC}.
\begin{table}
\setlength{\tabcolsep}{5pt}
\center
\begin{tabular}{c|c|c|c|c|c|c} 
\hline \hline 
Bin $[\rm GeV]$ &  \multicolumn{2}{c}{$\mathcal{R}_t$} & \multicolumn{2}{|c|}{$\mathcal{R}_b$} & \multicolumn{2}{c}{$\mathcal{R}_c$} \\
\hline
$[50,3000]$ & $-$ & $-$ & 0.99 & 0.99 & 0.99 & 0.99 \\
$[350,3000]$ & 0.97 & 0.97 & $-$ & $-$ & $-$ & $-$ \\
$[50,250]$ & $-$ & $-$ & 0.99 & 0.99 & 0.99 &0.99 \\
$[250,500]$ &  $-$ & $-$ & 1.00 &1.00 & 1.00 &1.00\\
$[350,500]$ &  0.98 &0.98 & $-$ & $-$ & $-$ & $-$\\
$[500,750]$ & 0.97 &0.97 & 0.99 &0.99  & 0.99 &0.99\\
$[750,1000]$ & 0.95 &0.95 & 0.98 &0.98 & 0.98 &0.98\\
$[1000,1500]$ & 0.94 &0.94 & 0.97 &0.97 & 0.96 &0.96\\
$[1500,2000]$ & 0.92 &0.92 & 0.95 &0.95 & 0.95 &0.95\\
$[2000,2500]$ & 0.90 &0.91 & 0.93 &0.94 & 0.93 &0.93\\
$[2500,3000]$ & 0.88 &0.89 & 0.92 &0.93 & 0.92 &0.92\\
$[3000,3500]$ & 0.87 &0.88 & 0.90 &0.91 & 0.91 &0.91\\
\hline \hline
\end{tabular}
\caption{The EWS corrections for heavy quark production at the LHC. The left (right) columns are for $\sqrt{s} = 7 \, (14) \, {\rm TeV}$.}
\label{table:tcorrectionsLHC}  \vspace{-0.45cm}
\end{table}

Preliminary measurements of the reconstructed $d \sigma/ d M_{t \bar{t}}$ spectrum have been reported by ATLAS~\cite{atlasnote} and CMS~\cite{boosted}, and
no large deviation from the SM has been found. At the large invariant masses studied in~\cite{boosted}, both the ${t  \bar{t}}$ production rate and the subtracted dijet background rate receive
large EWS corrections.\footnote{We thank Gilad Perez for discussion on this point.} Both  corrections act to increase the possibility for a non resonant excess of 
large $M_{t \bar{t}}$ events in this study,  since they suppress the SM rate, and should be taken into account before any precise conclusions can be drawn. A data driven normalization 
of the Monte Carlo estimation of the $d \sigma/ d M_{t  \bar{t}}$ spectrum, that is subsequently extrapolated to large $\hat s$ to search for
deviations in the tail from the SM expectation, is also susceptible to large EWS corrections.

As a specific example of the importance of these corrections, note that 
some plausible flavor symmetric models of new physics that could marginally explain the $A_{\rm FB}^{t\bar t}$ anomaly can cause a rise in the tail of $d \sigma/ d M_{t  \bar{t}}$ by
$\sim \!10\%$~\cite{Grinstein:2011yv}. Such an effect could be completely canceled by SM EWS corrections currently unaccounted for in MC simulation 
tools.

\begin{figure}
\includegraphics[bb=58 194 592 478,width=10cm]{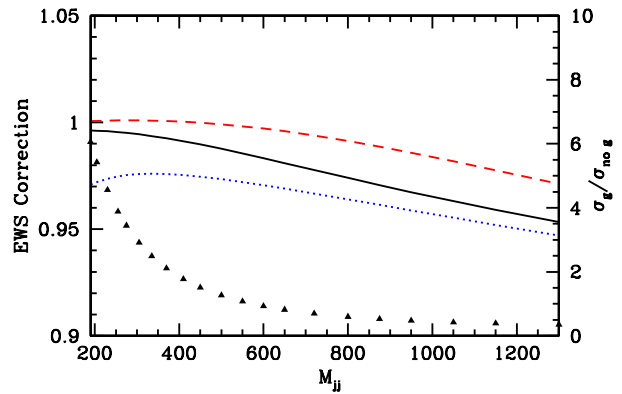}
\caption{EWS correction (left axis) to the Tevatron dijet spectrum as a function of dijet invariant mass (solid black). Also shown are the corrections to dijet processes involving external gluons (red dashed), and no external gluons (blue dotted). The black triangles are the ratio of cross sections (right axis) with and without  external gluons.}
\label{tevatroncorr}
\end{figure}

{\bf III. Dijets:}\label{dijets} EWS corrections are also important for dijet studies at the Tevatron and LHC.\footnote{Our results are consistent with previous results using infrared evolution equations~\cite{Fadin:1999bq} or a  $\rm SU(2)$ theory to sum Sudakov logarithms. Jets studies based on these techniques include~\cite{Kuhn:2001hz,Kuhn:1999nn,Kuhn:2009nf}. Fixed order EW corrections to dijet rates also give large corrections~\cite{Moretti:2005ut}. Our results are for the full $\rm SU(2) \times U(1)$ theory including $\gamma-Z$ mixing and Higgs effects.} 
We evaluate EWS corrections for all partonic
LO $2 \rightarrow 2$ QCD dijet processes with a rapidity cut, $|y|<1$, implemented as described in~\cite{Lane:1991qh} for the Tevatron dijet corrections.
The quark flavours ($u,c,s,d,b$) and gluon initial and final partonic states are summed over.
We average over the bin mass range, which is taken to be $10\%$ of the central value of the bin
as in~\cite{Aaltonen:2008dn}. The renomalization scale is $\mu = M_Z$. Varying the scale in the range
$\mu = (M_Z/2, 2 \, M_Z)$, or the choice of MSTW PDF eigenvalues used, changes the results by less than 1\%.
The EWS correction factor for the Tevatron is shown in Fig.~\ref{tevatroncorr}.

The total dijet rate involves partonic processes with and without external gluons, which cannot be separated experimentally.
The EWS corrections are very small for gluonic processes, since the gluon is an EW singlet.  This dilutes the overall importance of EWS effects for inclusive dijet production for low invariant mass events. To illustrate this, we have also shown in Fig.~\ref{tevatroncorr} the EWS corrections to dijet processes involving and not involving external gluons, as well as the ratio of these contributions to the total dijet rate. 
\begin{figure}
\includegraphics[bb=58 194 592 478,width=10cm]{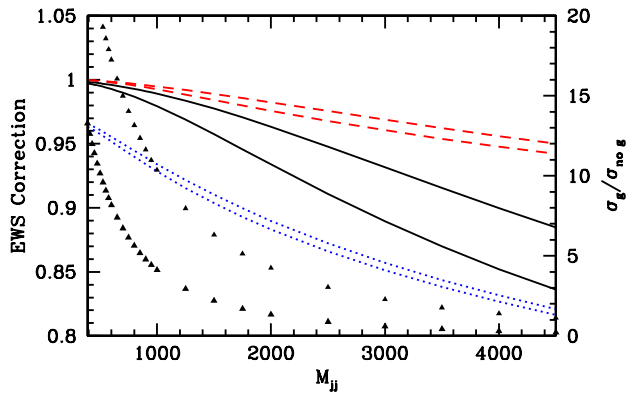}
\caption{Same as Fig.~\ref{tevatroncorr} for the LHC at $\sqrt{s} = 7 \, {\rm TeV}$ (lower curves) and 14~TeV (upper curves).}
\label{LHCcorr}
\end{figure}

The LHC results are shown in Fig.~\ref{LHCcorr}. We have imposed typical central jet rapidity cuts ($|y| < 2.8$) and cuts on the separation of the rapidity of the two leading jets  ($|\Delta y| < 1.2$) consistent with  the ATLAS study~\cite{atlas1}.
The results are insensitive to the particular rapidity cuts made. For example, varying the rapidity cut from $2.8 \rightarrow 2$ leads to a variation in the total EWS correction of less than $1\%$. We have also determined the EWS correction to the angular distribution measure $F_{\chi}[M_{jj}]$ as defined in~\cite{Aad:2011aj}.  EWS corrections suppress this ratio by $\sim 2 (5) \%$  for $2 (4) \rm {TeV}$ dijet masses, the correction factor to apply to the
SM calculation of $F_{\chi}[M_{jj}]$ is well approximated by $1 -0.128 M_{jj}^{0.258} + 0.143 M_{jj}^{0.239}$ for the mass range $0.5-5 \, {\rm TeV}$. This correction slightly relaxes angular distribution constraints on new physics.

We  find large corrections that must be included for precise studies of multi-TeV dijet events at both $\sqrt{s}=7$ and 14~TeV.
The importance of  EWS corrections in dijet studies changes with the LHC operating energy, since  the relative importance of gluonic
 and non-gluonic processes is largely driven by the PDF's. Gluonic dijet events become more important as the operating energy increases.
A recent ATLAS study~\cite{Collaboration:2011fc} has reported dijet events out to $M_{jj} \sim 5 \, {\rm TeV}$ with the 2011 data set, where the effects of EWS corrections
on the QCD partonic $2 \rightarrow 2$ dijet processes are significant, $\sim \!  - 20 \%$. 
These corrections can act to cancel a $t$-channel driven rise in the dijet invariant mass spectrum in models attempting to explain the $A_{\rm FB}^{t\bar t}$ excess that involve new light quark interactions. 

{\bf IV. Conclusions: \label{conc}} 
We have determined the EWS corrections for a number of observables of current interest.
EWS corrections enhance $A_{\rm FB}^{t \bar{t}}$ by a factor of $1.05$ and slightly reduce the tension between the SM prediction and the
measurement of $A_{\rm FB}^{t \bar{t}}(M_{t  \bar{t}} > 450 \, {\rm GeV})$ reported in~\cite{Aaltonen:2011kc}. They give a significant correction to the multi-TeV dijet spectrum at the LHC and are important for determining the tail of the dijet spectrum.
Many models constructed to explain $A_{\rm FB}^{t \bar{t}}$ introduce new interactions that increase the tail of this spectrum. This can be compensated for by the EWS corrections which are not included in MC simulation tools. Similarly, constraints based on the  extracted tail of the $d \sigma/ d M_{t\bar{t}}$ spectra are important when considering the $A_{\rm FB}^{t \bar t}$ anomaly; the SM prediction of the tail of this spectra
also receives large EWS corrections. 

{\bf Acknowledgements} 
We thank  Z.~Ligeti, B.~Pecjak, G.~Perez, G.~Rodrigo, J.~Rojo, P.~Skands, G.~Watt and B.~Webber for 
helpful communications. M.T. thanks the Aspen Center for Physics for hospitality while this work was in progress.


\end{document}